\documentclass[aps,prab,reprint,groupedaddress]{revtex4-2}
\usepackage{xcolor}
\usepackage{graphicx}
\usepackage{float}
\usepackage{amsmath}
\usepackage{amssymb}
\usepackage{dcolumn}
\usepackage{url}
\newcolumntype{.}{D{.}{.}{-1}}

\begin{document}


\title{Design, fabrication, and tuning of a THz-driven electron gun}


\author{Samantha M. Lewis}
\email{smlewis@fnal.gov}
\author{Julian Merrick}
\author{Mohamed A. K. Othman}
\author{Andrew Haase}
\author{Sami Tantawi}
\author{Emilio A. Nanni}
\affiliation{SLAC National Accelerator Laboratory}


\date{\today}

\begin{abstract}
We present the design, fabrication, and low power testing of a THz-driven field emission electron gun. The two cell standing-wave gun is designed to be powered by a 110 GHz gyrotron and produce 360 keV electrons with 500 kW of input power. Several gun structures were fabricated using a high precision diamond turned mandrel and copper electroforming. The field emission source is a copper tip with a 50~$\mu$m radius inserted halfway into first cell. The frequencies of the cavity resonances were mechanically tuned using azimuthal compression. This work presents electromagnetic and particle simulations of the design and cold test measurements of the fabricated structures.
\end{abstract}


\maketitle

\section{Introduction}
Next generation accelerator facilities require new electron source technology \cite{ElectronSources}. The properties of the electron source heavily influence the achievable beam parameters in applications such as Free Electron Lasers (FELs) and Ultrafast Electron Diffraction (UED). Existing facilities have proven extremely useful for studying novel materials and fundamental biological and chemical processes. To access new regimes of study, these systems require improvements in emittance, bunch length, energy spread, brightness, and synchronization. There are a number of research efforts dedicated to providing electron sources and acceleration techniques to meet these requirements.

A significant area of research is focused on increasing the accelerating gradient, which can help prevent beam degradation and provide high brightness sources in a small footprint. To date, advances towards high gradient acceleration have been made using a variety of techniques. The leading technology is Plasma Wakefield Acceleration (PWA), in which electrons are accelerated using wakes generated by a laser or particle beam \cite{PWAReview,PWAReview2,FACETPWA,PDPWA}. These experiments have demonstrated multi-GeV/m gradients and produced GeV-scale beams using staged acceleration. However, plasma instabilities pose challenges to delivering the beam performance required for future facilities.

In traditional normal conducting RF (NCRF) accelerator structures, the achievable gradient is limited by the breakdown voltage \cite{Kilpatrick}. The strength of the electric field in the cavity directly determines the accelerating gradient. Higher gradients require stronger electric fields, but vacuum breakdowns occur when the surface fields are too high. Experiments at various frequencies \cite{LimborgGun,AkreGun,BrownGun} have shown that the achievable surface electric field $E_s$ before breakdown increases with higher frequencies $f$ and shorter pulse lengths $\tau$. The achievable $E_s$ fundamentally limits the maximum possible gradient within commonly used NCRF cavities. A promising potential solution is developing similar accelerators at much higher frequencies.

Scaling accelerators to higher frequencies could provide performance benefits in addition to reaching higher gradients. Experiments have shown that breakdown behavior is influenced by the surface magnetic field and pulsed heating \cite{DolgashevBreakdown,LaurentPulsedHeating}. The timescale of pulsed heating is related to the pulse length $\tau$, which is in turn determined by the required fill time of the structure. High frequency structures require a shorter fill time, given by
\begin{equation}
t_F = \frac{2Q_L}{\omega} = \frac{2Q_0}{\left(1+\beta_c\right)\omega},
\label{eq:filltime}
\end{equation}
where $Q_L$ and $Q_0$ are the loaded and unloaded quality factors, $\beta_c$ is the coupling constant, and $\omega$ is the angular frequency. By reducing the required fill time and pulse length, pulsed heating can be reduced in high frequency cavities. Additionally, the shunt impedance scales as $f^{1/2}$ and the dissipated power per unit length scales as $f^{-1/2}$. These scalings could lead to lower losses and the ability to operate with high repetition rates. 

At the low end of the THz regime, it is possible to take advantage of frequency scaling improvements while still leveraging the extensive experience gained from operating traditional RF accelerators. Recent research efforts have made significant strides towards providing high power THz sources \cite{LCLSIITHz,SchmergeTHz,ORTHz,ZhongHPTHz}. The small size of THz cavities pose fabrication challenges, but also opportunities to use advanced fabrication techniques. Significant efforts have been made to study the fabrication and performance of copper THz structures, paving the way for the development of a high gradient THz accelerator \cite{LewisSplitBlock,NanniPrototyping,OthmanPrototyping}. This includes studies of split-block copper structures that have explored bonding techniques capable of withstand the high surface fields and provide the precision alignment needed for mm-scale cavities.

In addition to improvements from high gradient acceleration, there are performance requirements that necessitate new electron source technology. The quality of the electron source determines the achievable beam performance downstream. Generating and quickly accelerating short, low emittance electron bunches is fundamental to reaching new regimes of operation for both FEL and UED facilities. Electron source research efforts are underway to improve the performance of existing gun types as well as to develop new materials and techniques \cite{PierceMTE,VelardiNDphoto,PhotocathodeThesis,NicholsDFEA,Wimmer,LiJ,ShaoFE}. Electron guns can use a variety of emission mechanisms including thermionic, photo, and field emission. In the case of field emission, the surface electric field on the tip lowers the potential barrier such that electrons can tunnel from the cathode material, generating a beam. The high fields and fast timescales of THz cavities provide the ideal conditions to develop compact field emission and photo-field emission sources.

Combining these efforts, we have developed a two cell standing-wave electron gun at 110 GHz. The gun uses normal conducting copper cells that have been electroformed around a high precision mandrel. The electron source is a copper tip cathode located in the center of the first cell. At 110 GHz, copper can sustain multi-GV/m surface fields, which are sufficiently high to allow for field emission. The field-emitted bunch charge is expected to be 51 fC per RF cycle based on 3.9 GV/m fields on the copper cathode. The following sections present results of the electromagnetic and 3D particle simulations used in the design of the gun.

Several gun structures were fabricated, cold tested, and tuned. We detail the results of these cold tests and the mechanical tuning procedure. This represents the first demonstration of mechanical tuning of W-band accelerator cavities. This is also the first W-band accelerator structure made through electroforming. The use of electroforming allows for the cathode tip to be fabricated in one piece with the cells, which would not be possible with split block machining.

\section{Design and modeling}
\begin{figure}[!htb]
\centering
\includegraphics[width=0.8\columnwidth]{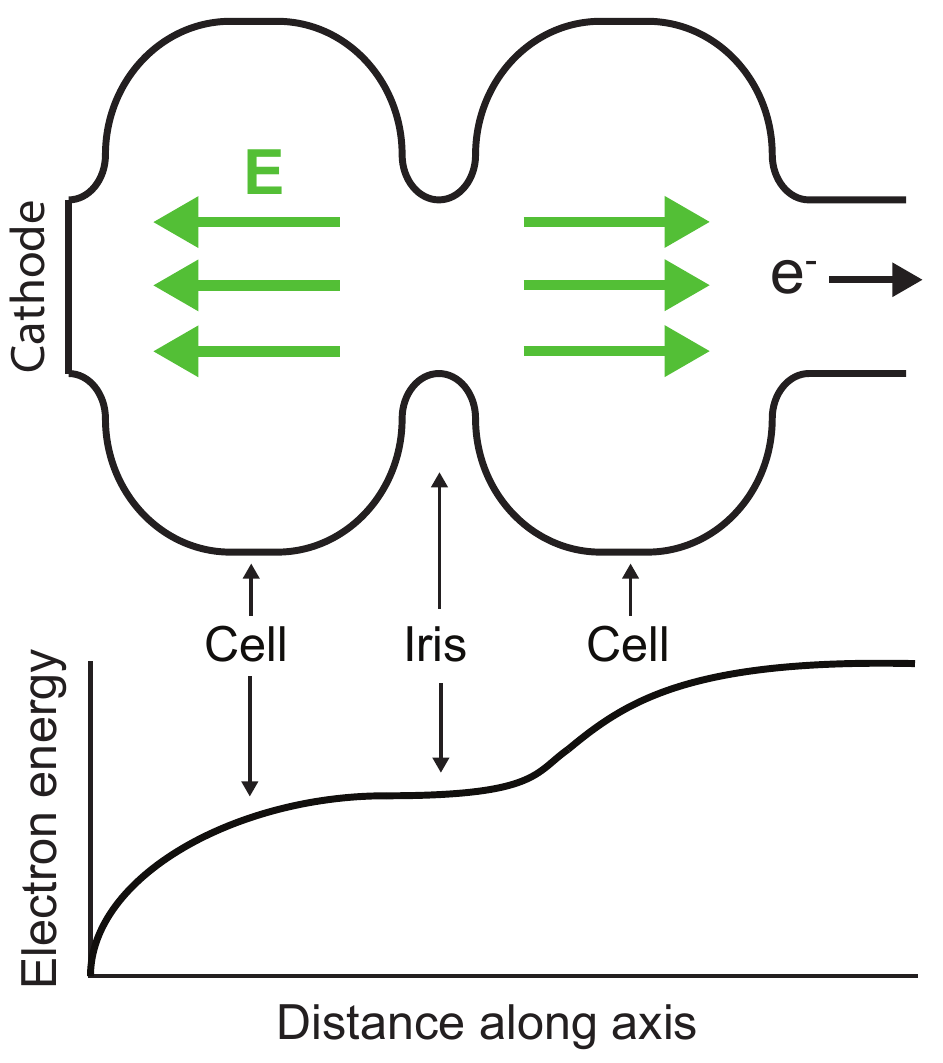}
\caption{Simplified schematic of a two cell gun, where electrons are accelerated from the cathode. The frequency and cell length are chosen to give continuous acceleration for a given operating mode. This schematic shows the $\pi$ mode where there is a 180 degree phase shift between cells.}
\label{fig:gunschematic}
\end{figure}

The electron gun is designed as a two cell, standing wave structure. In a standing wave electron gun, the electromagnetic field is excited by an external RF source. The electric field on-axis accelerates the beam from the cathode. The cell length and frequency are designed such that the phase shift between the cells gives consistent acceleration for a given electron beam energy. This is illustrated in the schematic in Figure \ref{fig:gunschematic}. The mode of operation shown is the $\pi$ mode, which has a 180 degree shift for each cell.

\begin{figure}[!htb]
\centering
\includegraphics[width=\columnwidth]{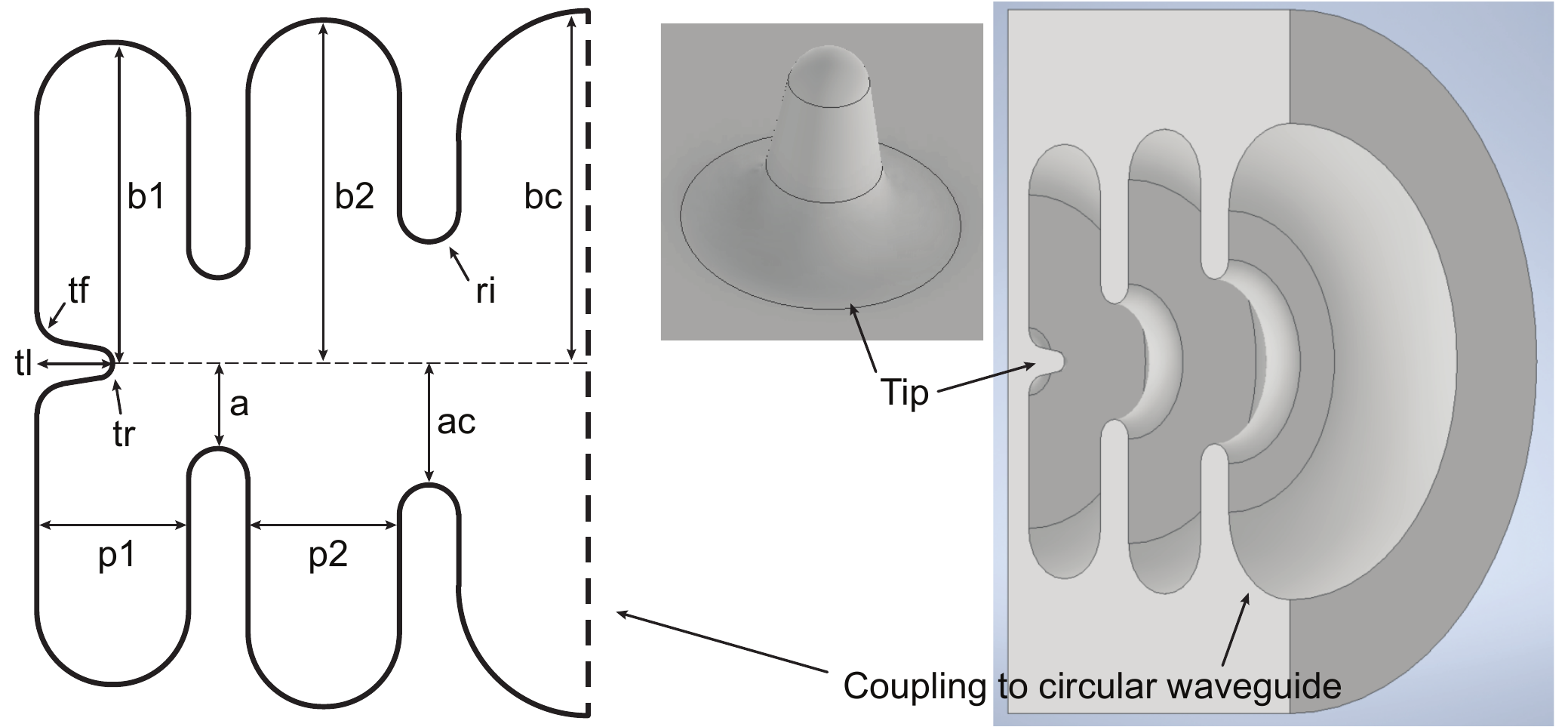}
\caption{Schematic and model of the copper tip gun. The structure is cylindrically symmetric around the horizontal dashed line shown. The parameters of the design are summarized in Table \ref{tab:cutipparams}. The structure consists of two cells, a copper tip (the cathode) located in the center of the first cell, and a coupling section. Input power is coupled in on-axis through a circular waveguide.}
\label{fig:cutipschematic}
\end{figure}

A schematic and model of the gun structure are shown in Figure \ref{fig:cutipschematic}. The design values of each parameter are summarized in Table \ref{tab:cutipparams}. A 50 $\mu$m radius copper tip serves as the field emission source. The electromagnetic performance of the gun cells was modeled using Ansys's High Frequency Structure Simulator (HFSS) \cite{HFSS}. The structure was designed such that the electric field is strongest in the first cell, providing high surface fields on the tip. For 500 kW of input power, the field on the tip was calculated to be 3.9 GV/m, which is roughly 4.5 times the highest surface field elsewhere in the structure.

\begin{table}[!htb]\centering
\begin{ruledtabular}
\begin{tabular}{c.l}
Parameter & \multicolumn{1}{c}{Value (mm)} & Description\\\hline
a & 0.286 & Iris radius\\
ac & 0.408 & Coupling iris radius\\
b1 & 1.080 & Radius of first cell\\
b2 & 1.155 & Radius of second cell\\
bc & 1.185 & Radius of input circular waveguide\\
p1 & 0.51 & Length of first cell\\
p2 & 0.51 & Length of second cell\\
ri & 0.1 & Iris radius of curvature\\
tf & 0.1 & Tip base radius of curvature\\
tl & 0.255 & Tip length\\
tr & 0.050 & Tip radius of curvature\\
\end{tabular}
\end{ruledtabular}
\caption{Design values of the copper tip gun.}
\label{tab:cutipparams}
\end{table}

\begin{figure}[!htb]
\centering
\includegraphics[width=\columnwidth]{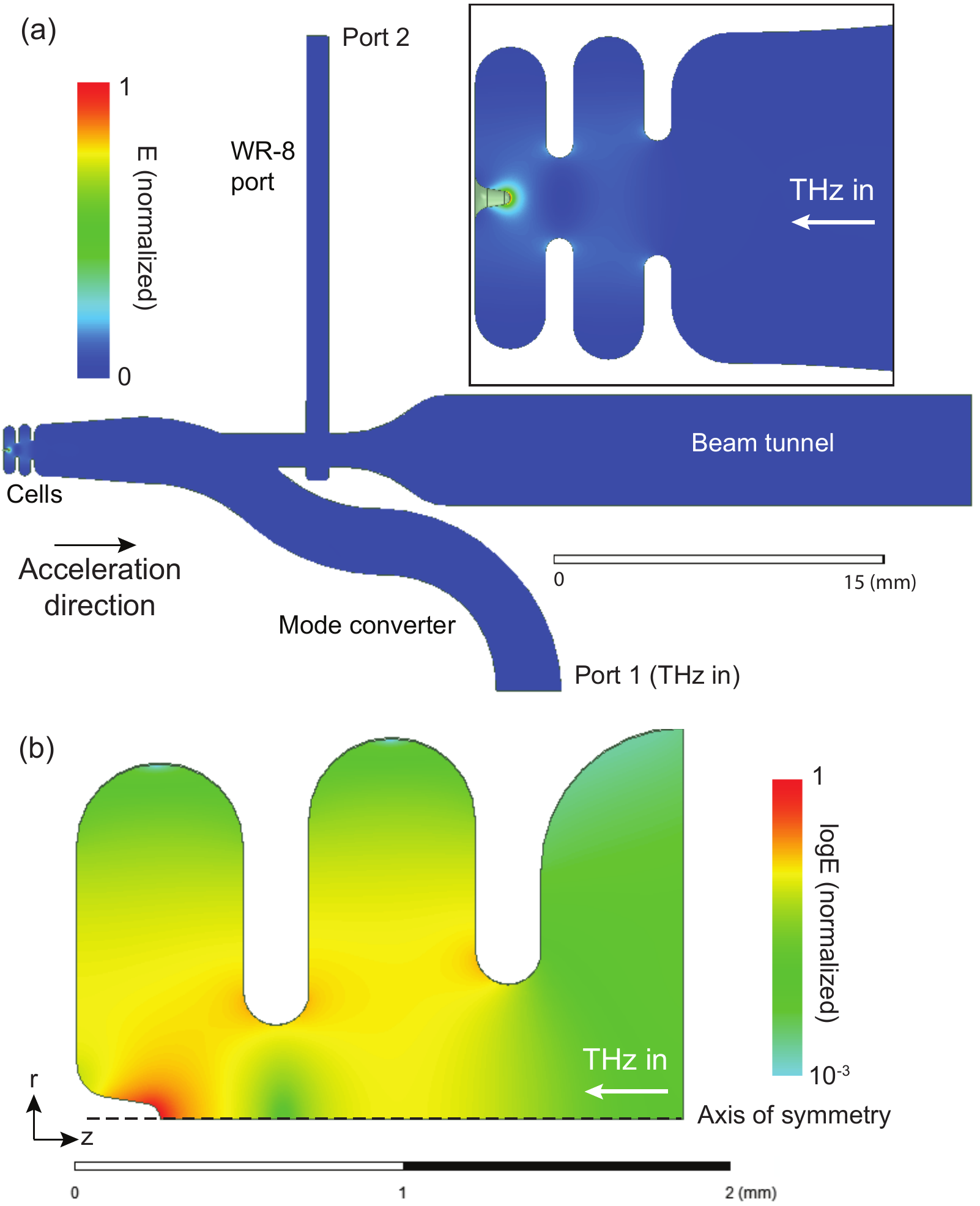}
\caption{(a) Electric field in the vacuum space of the mode converter and gun structure on the $\pi$ mode resonance. The simulation assumes mirror symmetry across the plane shown and uses copper surface resitivity for all external boundaries of the vacuum space excluding ports and the beam tunnel exit. The electric field is shown in normalized units. Power is input from port 1 and coupled into the cells. The inset shows the field profile in the cells including field enhancement at the tip. (b) Log-scale plot of the electric field in the cells, showing the field balance between the two cells. This simulation assumes cylindrical symmetry around the dashed line.}
\label{fig:MCHEmag}
\end{figure}

Power is supplied to the cells on-axis through a circular waveguide and mode converter. The gun was designed for use with a 110 GHz megawatt gyrotron which has been used to study breakdown in copper W-band cavities \cite{OthmanAPL}. The output beam from this gyrotron is transported in free space. To couple into the structure, the beam is focused onto a Gaussian horn which is followed by a mode converter to produce the TM$_{01}$ circular waveguide mode. This circular waveguide section of the mode converter mates with coupling section of the gun cells. Figure \ref{fig:MCHEmag} shows the mode converter and the simulated fields of the $\pi$ mode in the structure. The Gaussian horn and mode converter performance has been established by previous breakdown experiments \cite{NanniIPAC2016,NanniIPAC2018,OthmanAPL}. In this case, the mode converter wall has been modified to include a beam tunnel that does not affect the performance.

\begin{figure}[!htb]
\centering
\includegraphics[width=\columnwidth]{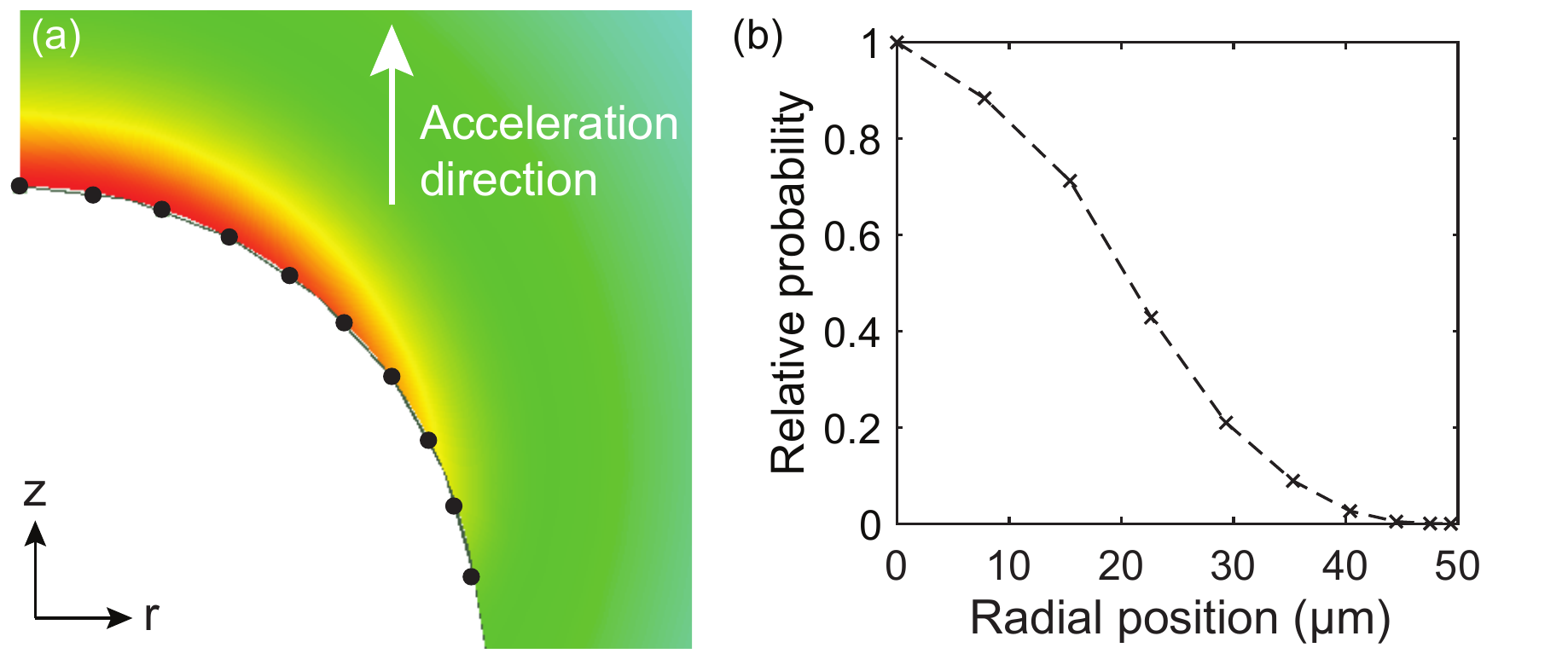}
\caption{(a) HFSS electric field on the tip showing the points which were used to discretize the emission area. (b) Relative probability of emission at those points based on the current density calculated from the Fowler-Nordheim equation.}
\label{fig:cutipemissionrad}
\end{figure}

\begin{figure}[!htb]
\centering
\includegraphics[width=\columnwidth]{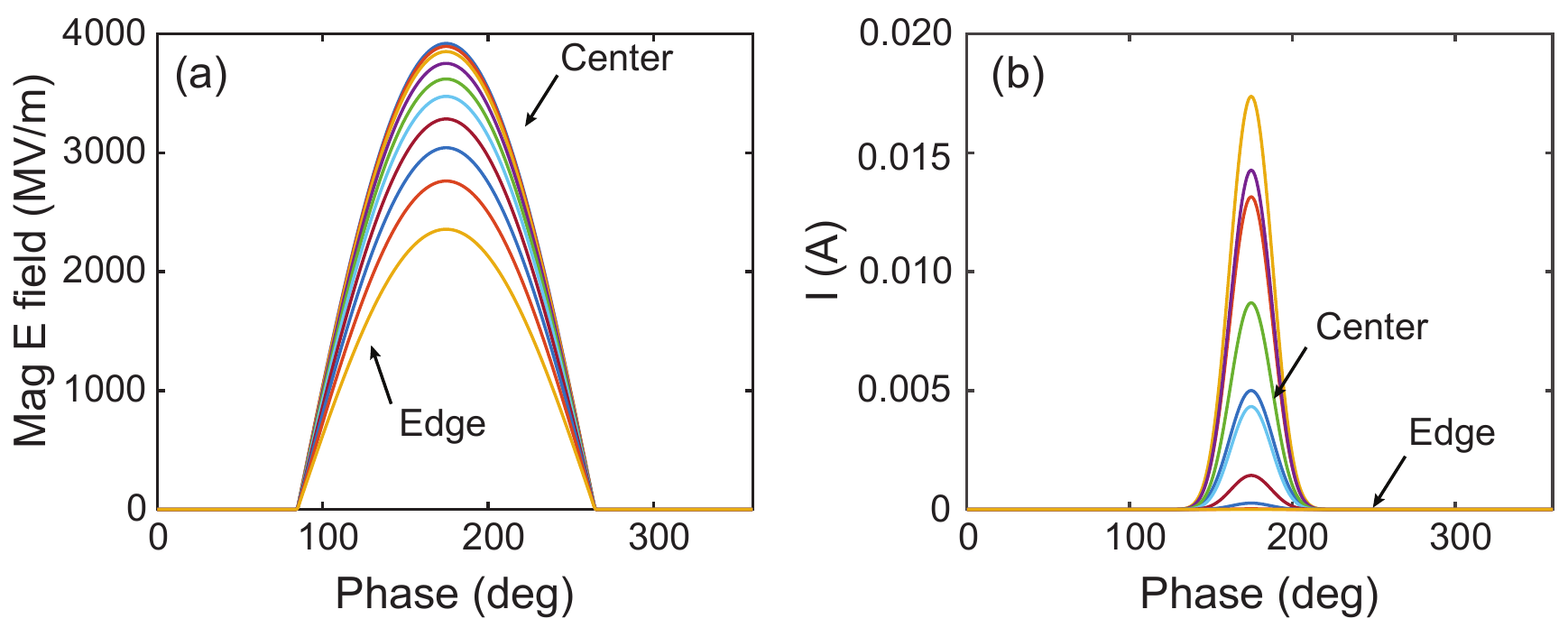}
\caption{(a) Magnitude of the electric field at the points shown in Figure \ref{fig:cutipemissionrad}. Half of the RF cycle is shown which is the correct orientation for acceleration. (b) The current versus phase based on the fields in (a). Most emission comes from the center and surrounding area. While the field is highest at the center of the tip, the current is higher at some points because they represent a larger area in a high-field region. The current density is highest at the tip center where the field is highest. The field is low enough at the edge of the hemisphere that it does not significantly contribute to the emission. This is reflected in the radial probability shown in Figure \ref{fig:cutipemissionrad}b.}
\label{fig:cutipemission}
\end{figure}

Particle simulations of the beam dynamics were performed using General Particle Tracer (GPT) \cite{GPT} combined with the HFSS fields and a Fowler-Nordheim-based model of the field emission. The Fowler-Nordheim equation describes cold field emission as a function of an applied electric field $E$ and a material-dependent work function $\phi$. The equation can be written as
\begin{equation}
\begin{split}
j_F =~&\frac{1.54\times10^{-6}\times10^{4.52\phi^{-0.5}}E^2}{\phi}\times\\
&\exp\left(-\frac{6.53\times10^9\phi^{1.5}}{E}\right)~\left[\text{A/m}^2\right]
\end{split}
\label{eq:jFN}
\end{equation}
with $E$ in units of V/m and $\phi$ in units of eV \cite{FowlerNordheim,WangThesis}. Current can be calculated for a given area $A_e$ including a field enhancement factor $\beta$. A simple model was developed to calculate the field-emitted bunch charge based on the simulated surface fields in HFSS. The copper tip was discretized using 10 points along the 50 $\mu$m radius. The field at each point was used to represent a ring of surface area on the tip hemisphere assuming azimuthal symmetry. The current $I$ of each of these rings was calculated based on the corresponding area. These points are shown in Figure \ref{fig:cutipemissionrad}. The resulting current versus RF phase is shown in Figure \ref{fig:cutipemission}. The current density is highest at the center, which is expected based on the field distribution. The field at the outermost points is low enough such that the emitted current is effectively zero. The total bunch charge emitted from the tip in one RF cycle was calculated to be 51 fC based on 500 kW of input power. No additional field enhancement factor was included. In reality, surface features will lead to additional local field enhancement. Thus, this calculation represents a minimum expected bunch charge for a smooth surface.

\begin{figure}[!htb]
\centering
\includegraphics[width=\columnwidth]{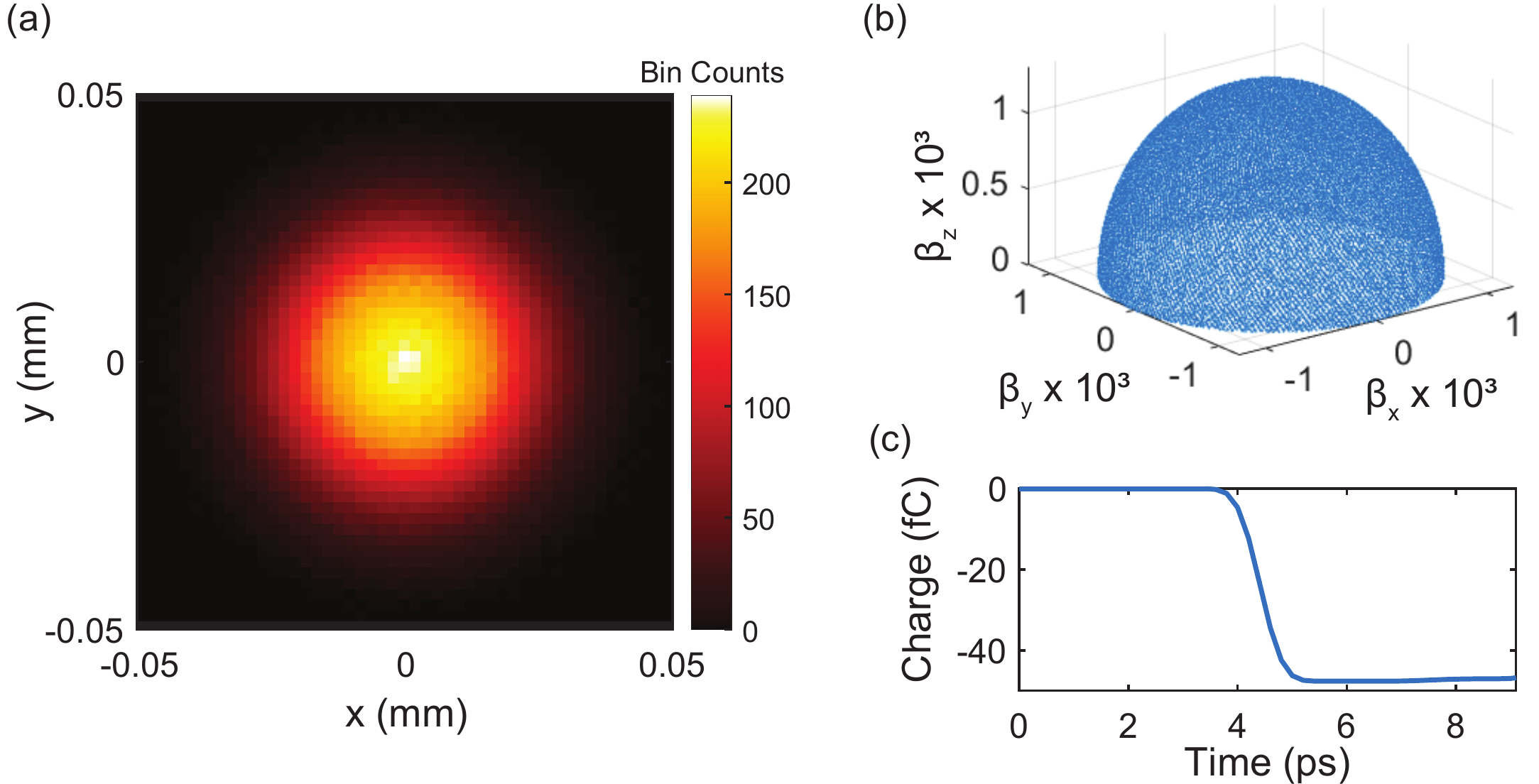}
\caption{Cumulative (a) spatial and (b) angular distribution of a beam. The radial distribution is set by the relative probability calculated using the HFSS fields on the tip. Emission is concentrated in the center and there is very little emission at the edges. (c) The build up of charge versus time based on the emission versus phase shown in Figure \ref{fig:cutipemission}. The maximum value is never completely reached because some particles are lost while others are being emitted. One RF cycle is shown.}
\label{fig:cuGPTinitialdist}
\end{figure}

To represent the emission in GPT, the total bunch charge was distributed in space and time based on the Fowler-Nordheim calculations. The current density versus radius was normalized and used to set the probability of the spatial distribution of the macroparticles. The model assumed symmetry in angle and a uniform $2\pi$ sr distribution in phase space. The probability of emission versus time was based on the emission versus phase at the tip center. All particles were assumed to be born at $z = 0$ with $E_0 = 0.4$ eV. The cumulative distribution of the emitted beam is shown in Figure \ref{fig:cuGPTinitialdist}. 

\begin{figure}[!htb]
\centering
\includegraphics[width=\columnwidth]{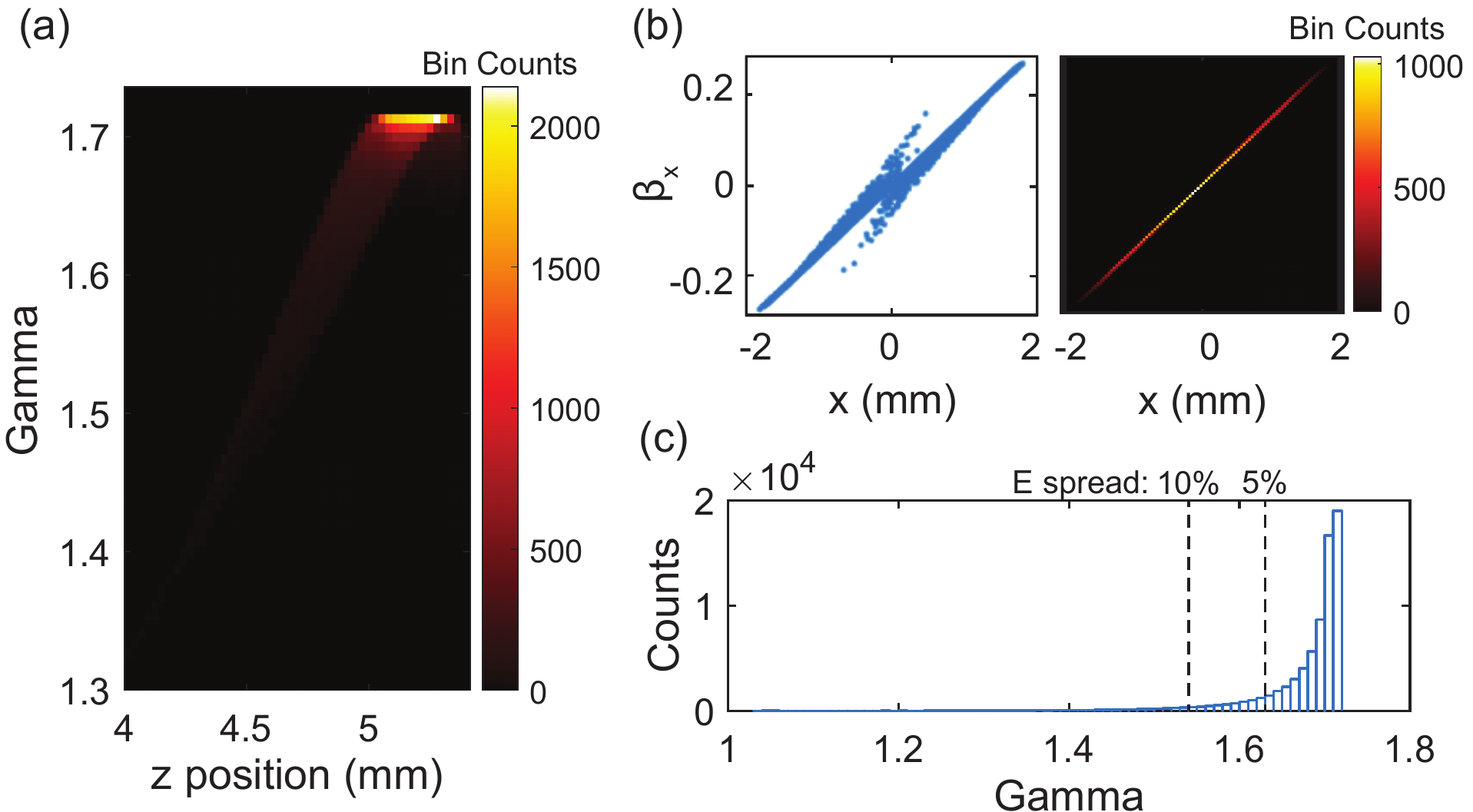}
\caption{(a) Longitudinal phase space after acceleration and a drift. The peak is highly localized near the maximum achievable energy. (b) Scatter and density plots of the transverse phase space after acceleration and a drift, showing the beam diverging. The beam is symmetric in $x$ and $y$ and thus only $x$-$\beta_x$ is shown. (c) Histogram of the electron energy distribution. The clear peak is at around $\gamma =$ 1.71--1.72 or about 365~keV, very close to the single particle prediction. The energy bins are $\gamma = 0.01$ or roughly 5 keV. The dashed lines indicate the cutoffs for 5\% and 10\% energy spread from the maximum value.}
\label{fig:cuGPTresults}
\end{figure}

To model the beam acceleration, the GPT simulations used the full 3D complex electric and magnetic fields of the gun structure calculated in HFSS for 500 kW of input power. Space charge was included using a built-in 3D mesh-based routine. The beam energy and phase space after acceleration and a drift are shown in Figure~\ref{fig:cuGPTresults}. The maximum kinetic energy is 366 keV, corresponding to $\gamma = 1.717$. While the core of the bunch is accelerated, there is a low energy tail. Roughly 74\% of the particles fall within 5\% of the maximum energy, and 85\% fall within 10\% of $E_{max}$. The normalized transverse rms emittance was calculated using the built-in GPT routine. The emittance for particles within 5\% of the peak energy is $\epsilon_{N} = 0.88~\text{mm-mrad}$ and for particles within 10\% of the peak it is $\epsilon_{N} = 1.04~\text{mm-mrad}$.  These values are on par with existing photocathode values and are the expected order of magnitude for a 50 $\mu$m radius tip.

The GPT simulations included modeling of the full beam transport for high power testing. The nominal operating point for the design of the assembly was chosen to be 500 kW of input power, and simulations were performed to confirm that the gun could provide acceleration for input powers as low as 50 kW. Discussion of the design of the magnets and beam diagnostics for high power testing will be presented in a future paper.

\section{Cold tests and tuning}
\begin{figure}[!htb]
\centering
\includegraphics[width=\columnwidth]{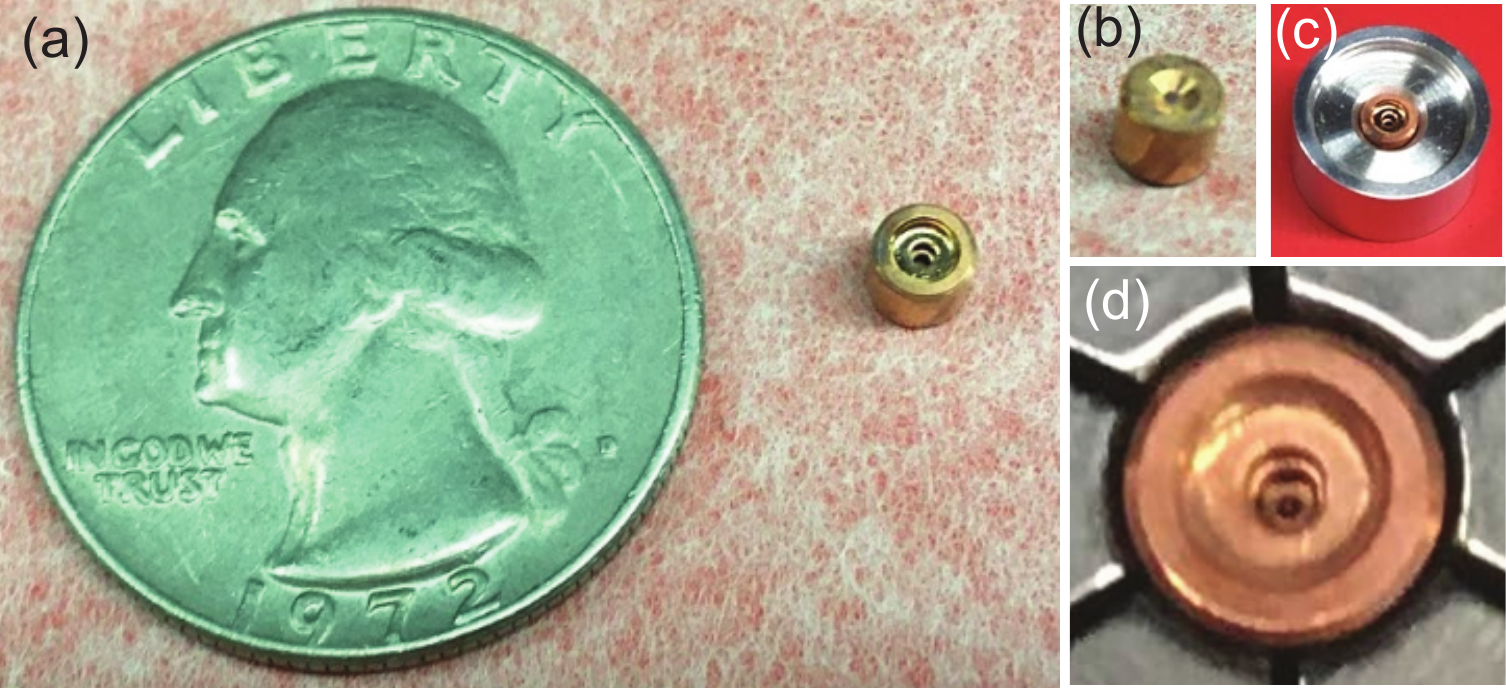}
\caption{Photos of the fabricated gun structures. (a) Cells with the coupling opening pointed up. The coupling iris and inner iris are both visible. (b) Back of the structure showing a divot for centering and alignment in the final assembly. (c) Cell structure upright in an aluminum adapter for cold testing. (d) Zoomed photo where inner features are visible. The copper tip is visible in the center and the inner iris is also visible.}
\label{fig:ctphotos}
\end{figure}

Multiple gun structures were fabricated via electroforming. The vacuum space of the structures was machined into an aluminum mandrel using diamond turning. The tip portion was removed from the mandrel using electrical discharge machining (EDM). Copper was electroformed around the aluminum mandrel and the mandrel was dissolved using sodium hydroxide. The electroformed structure includes the cells, tip, and coupling section as one monolithic piece. Photos of one structure are shown in Figure \ref{fig:ctphotos}.

\begin{figure}[!htb]
\centering
\includegraphics[width=\columnwidth]{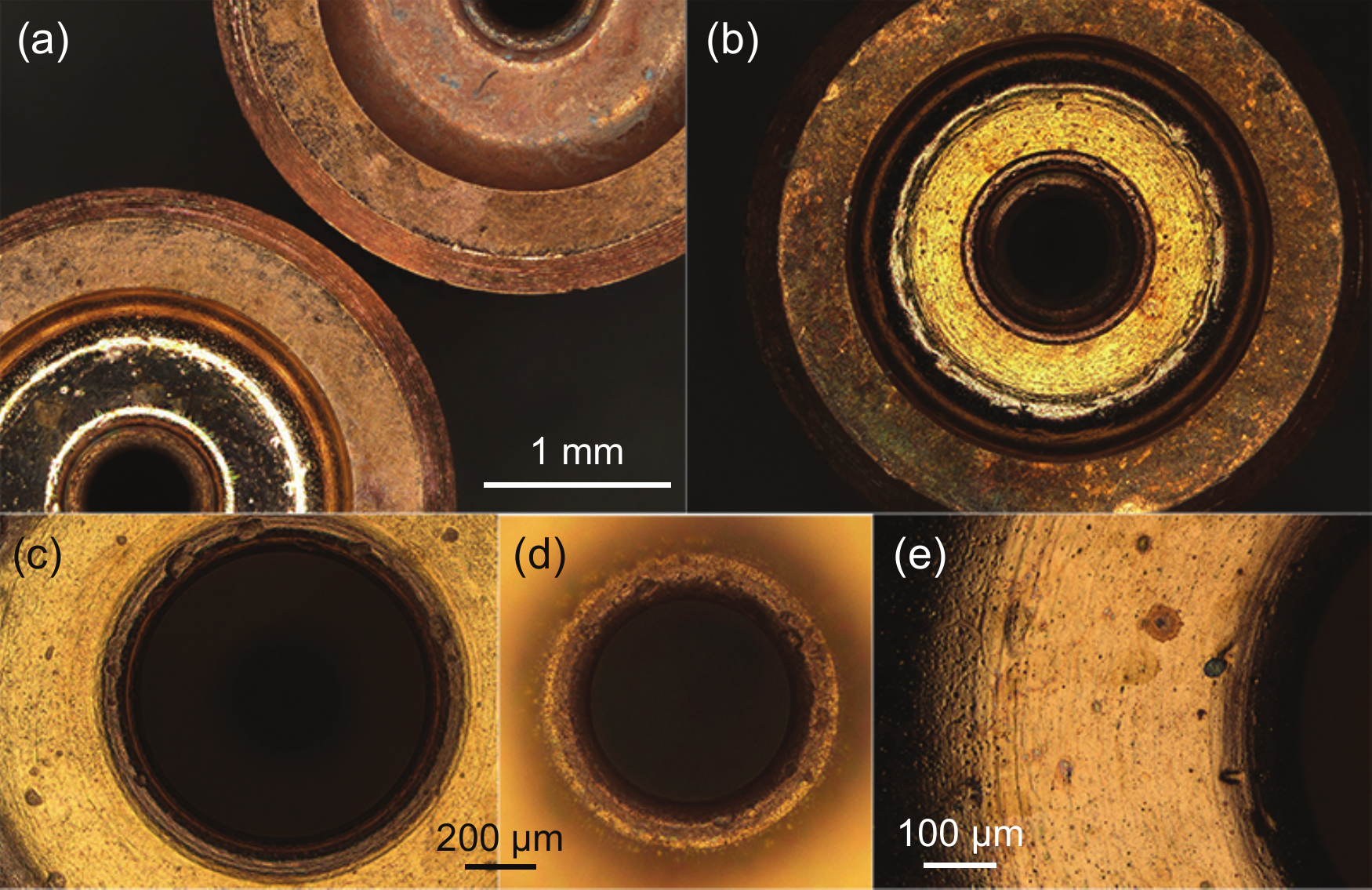}
\caption{Microscope images of different gun structures. All of the images are stacked focus images. (a) Side-by-side comparison of two structures showing the different surface finishes. (b) Image of a structure showing both irises and the input coupling section. The surface finish on the outer surface which mates with the input waveguide shows more wear-and-tear than the cell surfaces. (c) Zoomed view of the coupling iris. There are pits visible in the iris. These features will likely be smoothed during high power processing. (d) The inner iris, viewed with the same magnification as (c). (e) Close-up view of the coupling iris edge and surface of the coupling cell.}
\label{fig:ctmicroscopeimages}
\end{figure}

\begin{figure}[!htb]
\centering
\includegraphics[width=\columnwidth]{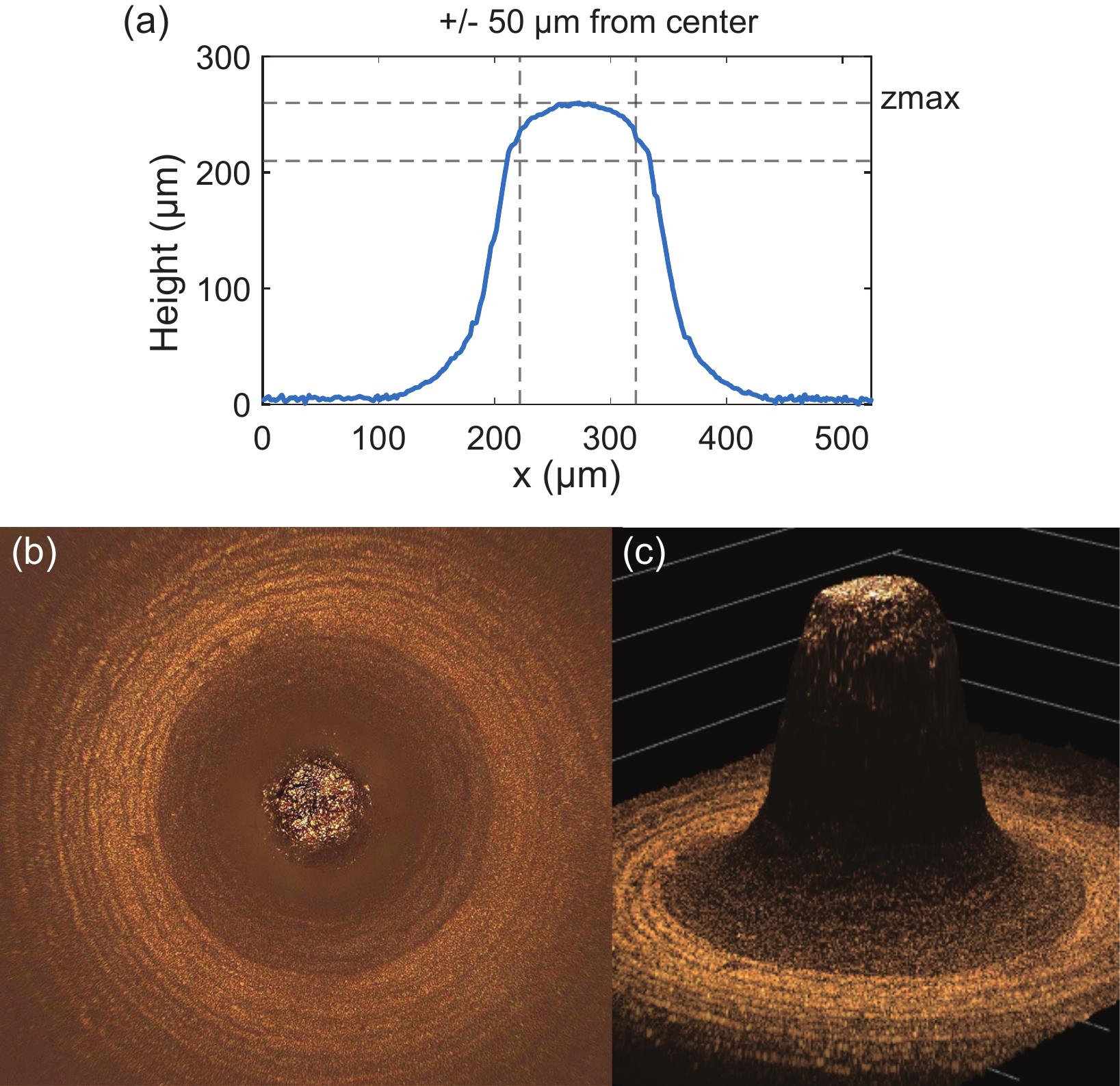}
\caption{(a) Height profile through the center of a copper tip. The dashed lines indicate the height and the distance corresponding to a 50 $\mu$m radius, the design value. The tip is not a perfect hemisphere, but its overall size matches well with the design. The height from the baseline is 255 $\mu$m, which was the design value for the height. (b) Top-down view of a tip in a stacked-focus image. (c) Height data displayed in 3D form.}
\label{fig:tipimages}
\end{figure}

A laser confocal microscope was used to image each structure in detail. Images of several structures are shown in Figure \ref{fig:ctmicroscopeimages}. The microscope was used to measure the 3D profile of the cathode tip. Example height data and 3D images are shown in Figure \ref{fig:tipimages}. Overall the measured parameters of the fabricated structures match well with the design, with small variations across the different structures.

\begin{figure}[!htb]
\centering
\includegraphics[width=\columnwidth]{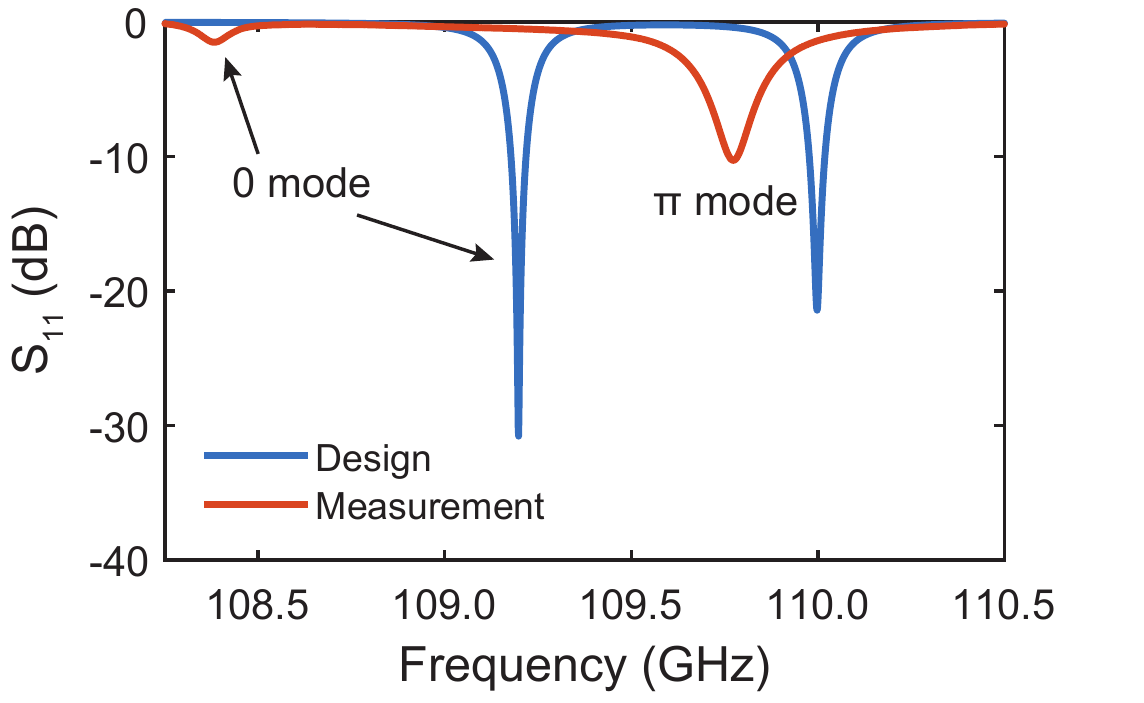}
\caption{An example measurement of the reflection signal $\left(S_{11}\right)$ compared to the design. The 0 mode and $\pi$ mode are indicated. In the fabricated structures, both modes are lower in frequency than the design. There is also a clear difference in the relative strength of the two modes and their spacing.}
\label{fig:S11plot}
\end{figure}

Cold test measurements were performed on all of the fabricated structures. In each case, the cavity resonances were lower than the design values of 109.22~GHz and 110.01~GHz for the 0 and $\pi$ modes, respectively. The range of measured 0 mode frequencies was 108.17--108.54~GHz, and the $\pi$ mode frequencies ranged over 109.61--109.90~GHz. The relative strength of the two modes and the mode spacing also differed significantly from the design. A comparison is shown in Figure \ref{fig:S11plot} with a representative measurement. 

Additional modeling was performed to determine potential causes of the variation and investigate the resulting impact on the electron beam. The frequency shifts could not be explained by the structure being uniformly oversized. Instead, the simulations showed that there is a difference in the fabrication error of the radii of the two cells (b1 and b2), altering the field balance. Based on measurements of the mandrels before electroforming, it is also likely that there was an error in the length of the first cell. The full details of this study are beyond the scope of this paper, but can be found in \cite{LewisThesis}. The frequency shifts can be explained by deviations in these dimensions in the range of 1--10~$\mu$m.

The operational tuning range of the gyrotron source is 110.08--110.1 GHz. The $\pi$ mode did not fall within this range in any of the structures, making it necessary to tune the cells. Further, HFSS simulations matching the measured spectrum had a weaker electric field in the first cell and a stronger field in the second cell. GPT simulations based on these fields showed that the structure would still provide acceleration with a lower maximum energy. The reduced field in the first cell would also significantly lower the field on the tip, limiting the achievable bunch charge. Several methods of tuning were attempted to raise the frequency and correct the field balance. Active thermal tuning alone would be insufficient to bring the frequency within range for high power operation. Instead, a combination of chemical etching and mechanical tuning was required. Etching using HCl acid raised the frequency of both modes in all of the structures that were etched. This indicates there was some residual material---potentially aluminum from the mandrel---that affected the frequency.

\begin{figure}[!htb]
\centering
\includegraphics[width=\columnwidth]{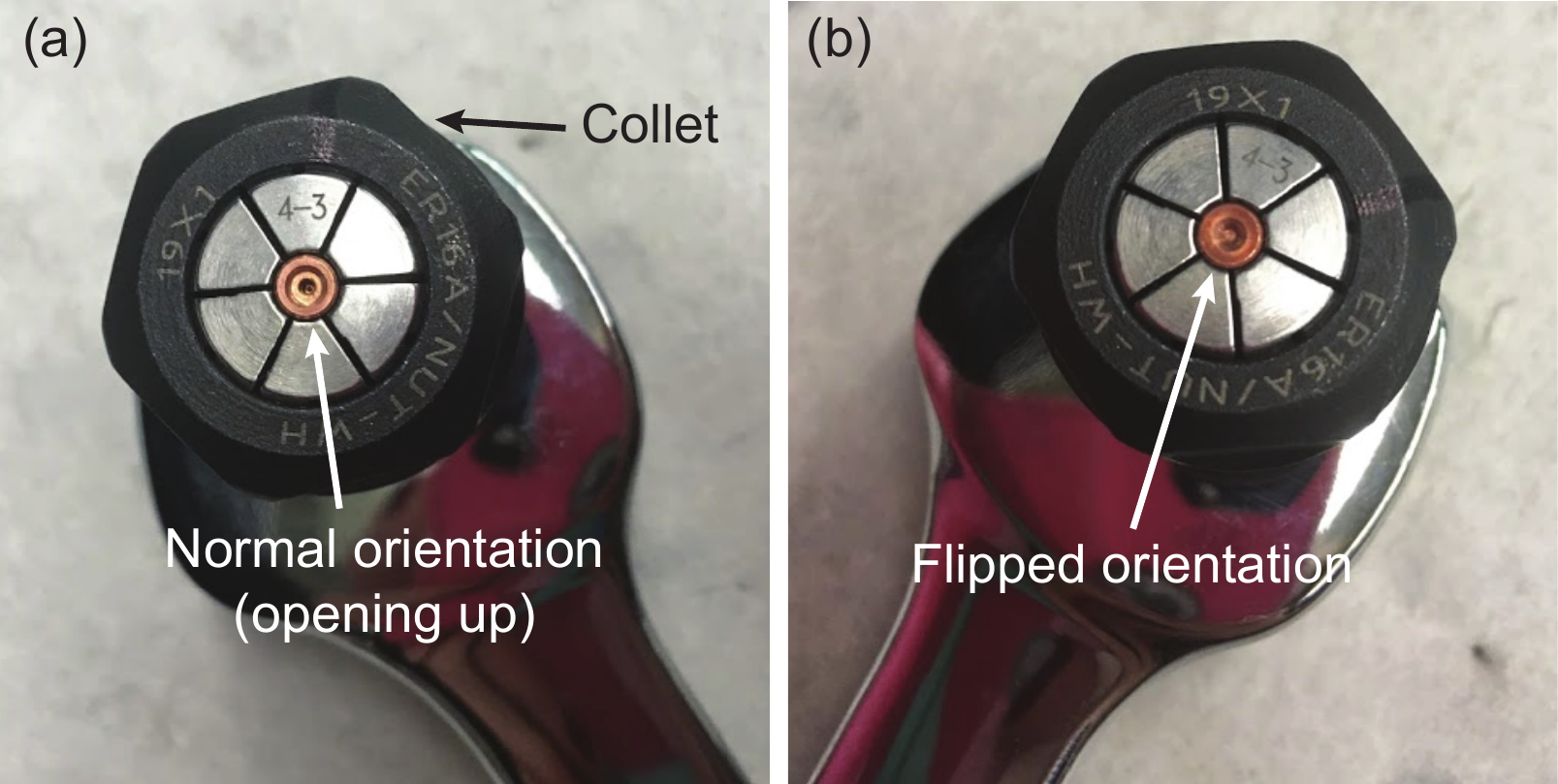}
\caption{Images of a structure in the tuning collet. (a) The `normal' orientation with the circular input waveguide opening pointing up. (b) The `flipped' orientation.}
\label{fig:mechtuningphoto}
\end{figure}

\begin{figure}[!htb]
\centering
\includegraphics[width=\columnwidth]{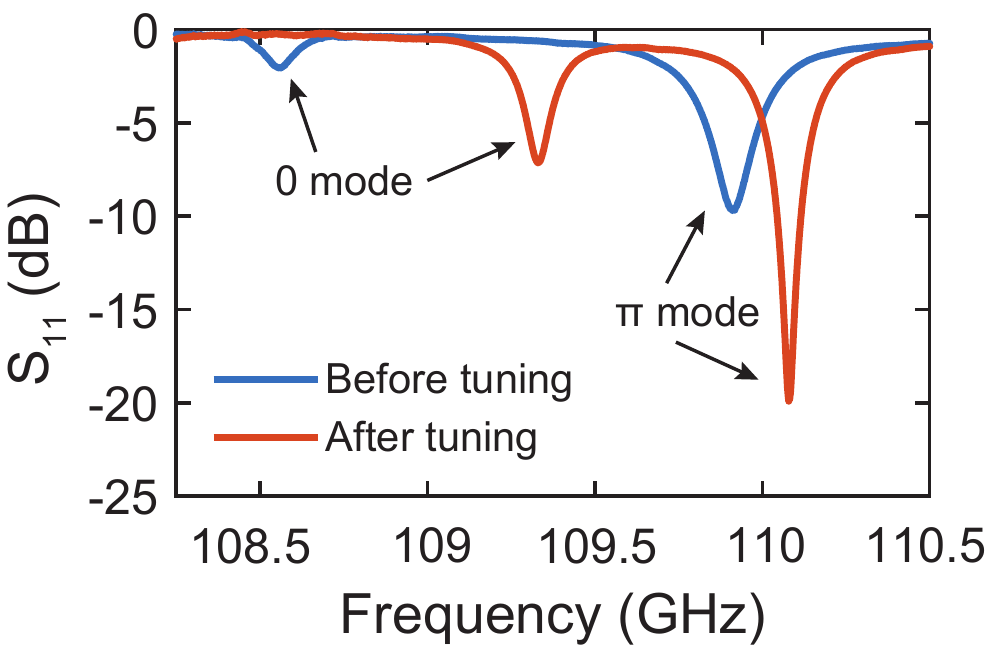}
\caption{The measured reflection $S_{11}$ before and after tuning. The 0 mode and $\pi$ mode are indicated. The 0 mode was intentionally tuned by a larger degree than the $\pi$ mode in order to achieve better field balance for acceleration and emission.}
\label{fig:c4tuningres}
\end{figure}

Following etching, several structures were mechanically tuned using an ER collet to achieve azimuthal compression. Photos of a gun structure in the collet during tuning are shown in Figure \ref{fig:mechtuningphoto}. The orientation of the structure in the collet affected the degree of tuning of the two modes. Using a careful series of small tuning steps, it was possible to tune the 0 mode by a larger shift than the $\pi$ mode. The resulting spectrum matches more closely with the original design. Additional modeling was performed to match the measured tuned result. These simulations indicate the field balance and surface field on the tip should provide significantly better performance than the structures as-fabricated. The measured $S_{11}$ of the final structure before and after mechanical tuning is shown in Figure \ref{fig:c4tuningres}. The final frequency of the $\pi$ mode was 110.081 GHz. Measurements under vacuum showed a shift to the expected value of roughly 110.114 GHz, which was measured in multiple air to vacuum cycles.

\section{Summary and future work}
We have developed a field emission electron gun at 110~GHz. The gun will be operated at room temperature and is expected to produce 51 fC, 360 keV electron bunches with 500 kW of input power. The gun cells were fabricated using electroforming, which provided a high degree of accuracy on observable surfaces. Small deviations on the order of 1--10 $\mu$m were able to be corrected using etching and mechanical tuning.

The gun cells are part of a larger assembly which will be used to characterize the beam during high power testing. In addition to measuring the energy spread, beam size, and field-emitted current, the high power measurements will include breakdown monitoring and build on previous W-band breakdown studies. Details on the assembly and beam characterization setup can be found in \cite{LewisThesis}. The results of the high power measurements will be reported in a future paper. The gun is capable of providing acceleration over a range of input powers and thus the emission properties of the cathode will be examined at multiple operating points. The gun cells are fully demountable from the rest of the assembly and multiple structures have been tuned within the gyrotron's operational range. This will provide the opportunity to study the efficacy of different tuning methods and provide confirmation of the effects of the mode balance on acceleration and emission.

Future work will include multiple stages of W-band acceleration after the gun to develop a THz-driven, normal conducting linac. Other cathode materials and types will also be explored, including the use of Diamond Field Emitter Arrays (DFEAs) and photo-field sources \cite{DFEASimakov,LewisIPAC2019,LewisThesis}.

\section{Acknowledgments}
We gratefully acknowledge Dale Miller, Andy Nguyen, Mario Cardoso, Walter Brown, and the SLAC TID Advanced Prototyping and Fabrication team for their assistance with structure fabrication. We would like to thank Dennis Palmer for his assistance with RF measurements and Emma Snively for her help with GPT simulations. We also thank our colleagues Julian Picard, Samuel Schaub, and Richard Temkin for valuable discussions about the gyrotron source and high power testing. This work was supported by the Department of Energy Contract No. DE-AC02-76SF00515 (SLAC) and by NSF Grant No. PHY-1734015.

\bibliography{draftbib}

\end{document}